\pdfoutput=1
\documentclass[preprint,journal]{vgtc}       

\ifpdf
  \pdfoutput=1\relax                   
  \usepackage{graphicx}                
  \DeclareGraphicsExtensions{.pdf,.png,.jpg,.jpeg} 
\else
  \usepackage{graphicx}                
  \DeclareGraphicsExtensions{.eps}     
\fi%

\graphicspath{{figures/}{pictures/}{images/}{./}} 

\usepackage{amsmath}
\usepackage{url}
\usepackage{microtype}                 
\PassOptionsToPackage{warn}{textcomp}  
\usepackage{textcomp}                  
\usepackage{mathptmx}                  
\usepackage{times}                     
\usepackage{cite}                      
\usepackage{tabu}                      
\usepackage{booktabs}                  

\onlineid{1069}

\vgtccategory{Research}
\vgtcpapertype{algorithm/technique}

\title{SemanticAxis: Exploring Multi-attribute Data by\ Semantics Construction and Ranking Analysis}

\author{Zeyu Li, Changhong Zhang, Yi Zhang and Jiawan Zhang, \textit{Senior Member, IEEE}}
\authorfooter{
\item
 Zeyu Li, Changhong Zhang, Yi Zhang and Jiawan Zhang are with the School of Computer Software, Tianjin University. E-mail: {lzytianda, ch\_zhang, yizhang, jwzhang}@tju.edu.cn.
\item
Changhong Zhang is with the Computer College of Qinghai Nationalities University.
}

\shortauthortitle{Li \MakeLowercase{\textit{et al.}}: SemanticAxis：exploring multi-attribute data by 
semantics construction and ranking analysis}

\abstract{Mining the distribution of features and sorting items by combined attributes are two common tasks in exploring and understanding multi-attribute (or multivariate) data. Up to now, few have pointed out the possibility of merging these two tasks into a united exploration context and the potential benefits of doing so. In this paper, we present SemanticAxis, a technique that achieves this goal by enabling analysts to build a semantic vector in two-dimensional space interactively. Essentially, the semantic vector is a linear combination of the original attributes. It can be used to represent and explain abstract concepts implied in local (outliers, clusters) or global (general pattern) features of reduced space, as well as serving as a ranking metric for its defined concepts. In order to validate the significance of combining the above two tasks in multi-attribute data analysis, we design and implement a visual analysis system, in which several interactive components cooperate with SemanticAxis seamlessly and expand its capacity to handle complex scenarios. We prove the effectiveness of our system and the SemanticAxis technique via two practical cases.} 


\CCScatlist{ 
 \CCScat{K.6.1}{Management of Computing and Information Systems}%
{Project and People Management}{Life Cycle};
 \CCScat{K.7.m}{The Computing Profession}{Miscellaneous}{Ethics}
}

\teaser{
  \centering
  \includegraphics[scale=0.485]{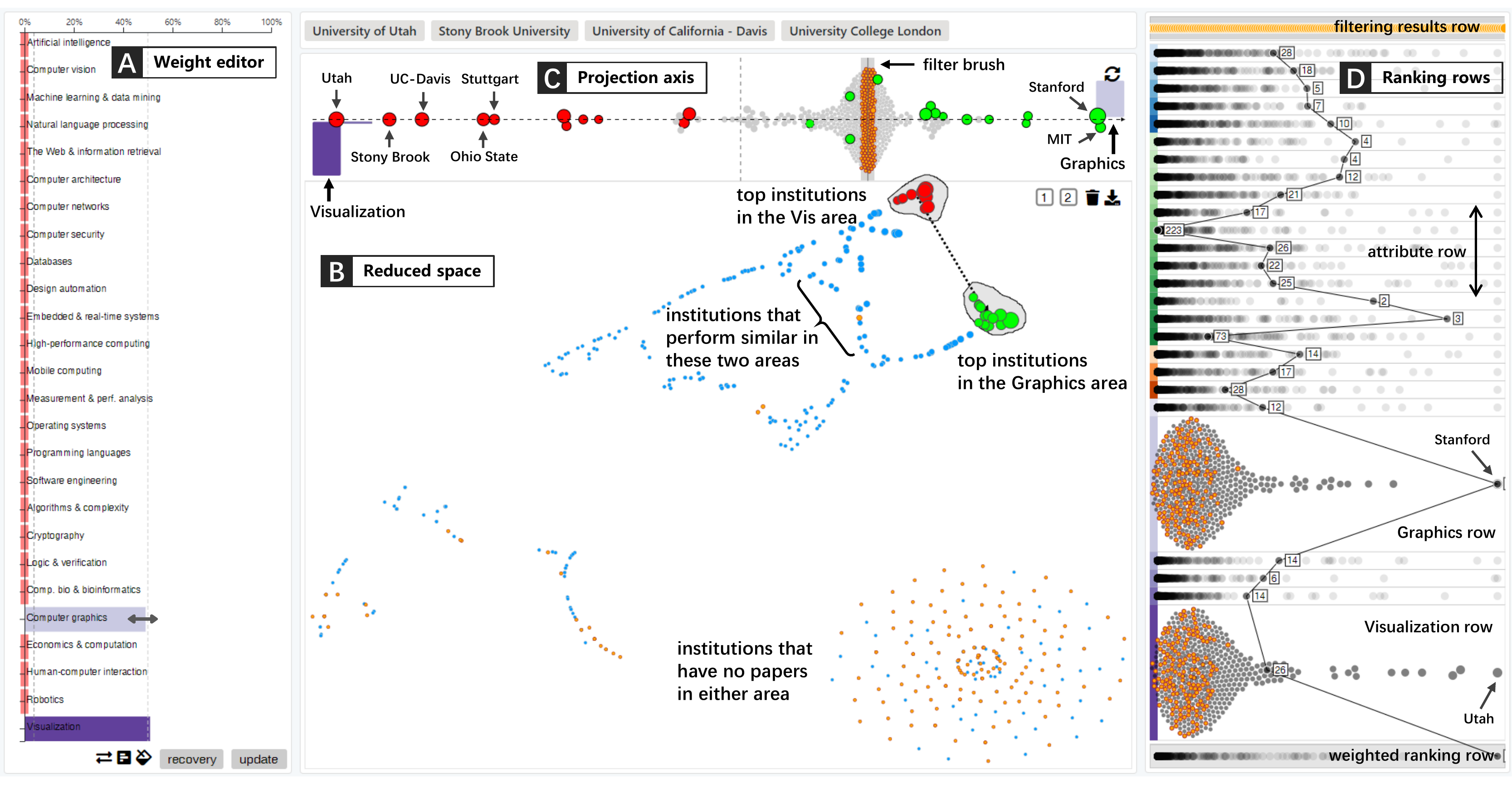}
  \caption{Interface of our system. With weight editor, analysts are free to emphasis or ignore attributes by adjusting their weights (the light red rectangle reminds the analyst that the weight of the corresponding attribute has been or will soon be reduced to zero). Reduced space presents dimension reduction results and allows analysts to construct SemanticAxis by lassoing two groups of points. Projection axis is designed for interpreting a constructed axis and checking its semantics distribution in reduced space. Ranking rows are used to verify inferences by providing details of data points (space for those emphasised attributes is expanded to enable individual checking).}
	\label{fig:interface}
}

\vgtcinsertpkg

\begin{document}


\firstsection{Introduction}

\maketitle

Multi-attribute (or multivariate) data is widely available in the real world and often rich in information. Hence, exploratory analysis towards understanding and interpreting multi-attribute data has always been a hot topic in visual analysis. Among existing work, dimension reduction (DR) is a commonly used technique. It can visually reveal the overall distribution pattern (e.g., a direction with certain semantics) and local characteristics of data (e.g., clusters and outliers). However, we believe that the following two points limit its application:

\begin{itemize}
  \item Sometimes it is difficult to understand DR results. Especially when we are unfamiliar with the data, there is no significant clusters in reduced space\footnote{Reduced space refers to the 2D plane created by DR algorithms.}, or data distribution does not match what it looks like in analysts' minds. Analysts often ask: What does this cluster mean? Why is there an outlier here? Is there a certain direction that may convey explicit semantics?
  \item It does not support data filtering and ranking based on single or combined attributes. However, these tasks are important and common in multi-attribute data analysis. For example, students expect to find universities that match their interests according to their performance in various research areas; teachers hope to ascertain students who are partial to several subjects according to their scores in each subject.
\end{itemize}

At present, few work intents to overcome these two limitations at the same time and points out the potential connections between them. In this paper, we present SemanticAxis, a technique that treats these two limitations as tasks and seamlessly merges them into a united exploration context. This is achieved by enabling analysts to interactively build semantic vectors that represent abstract concepts. To be specific, in reduced space, analysts can select the region of interest and another group of points as a target group and a control group, respectively. Then we compute the high-dimensional vector that connects the center of these two groups. Specially, if the vector has a distinct greater absolute value in some dimensions than others, and the combination of these dimensions can be interpreted as a reasonable concept by analysts, we call this vector a semantic axis. For instance, consider a multi-attribute data that describe scores of plenty of students across multiple subjects, and we have chosen two groups of students, who excel in natural sciences (e.g., mathematics and physics) and humanities (e.g., history and politics), respectively. If these selected students show no significant difference in other subjects, then the semantics of one end of the constructed semantic axis denotes ``they preform better in natural sciences than humanities" and the other end shows the opposite.

\begin{figure}
  \centering
  \includegraphics[scale=0.4]{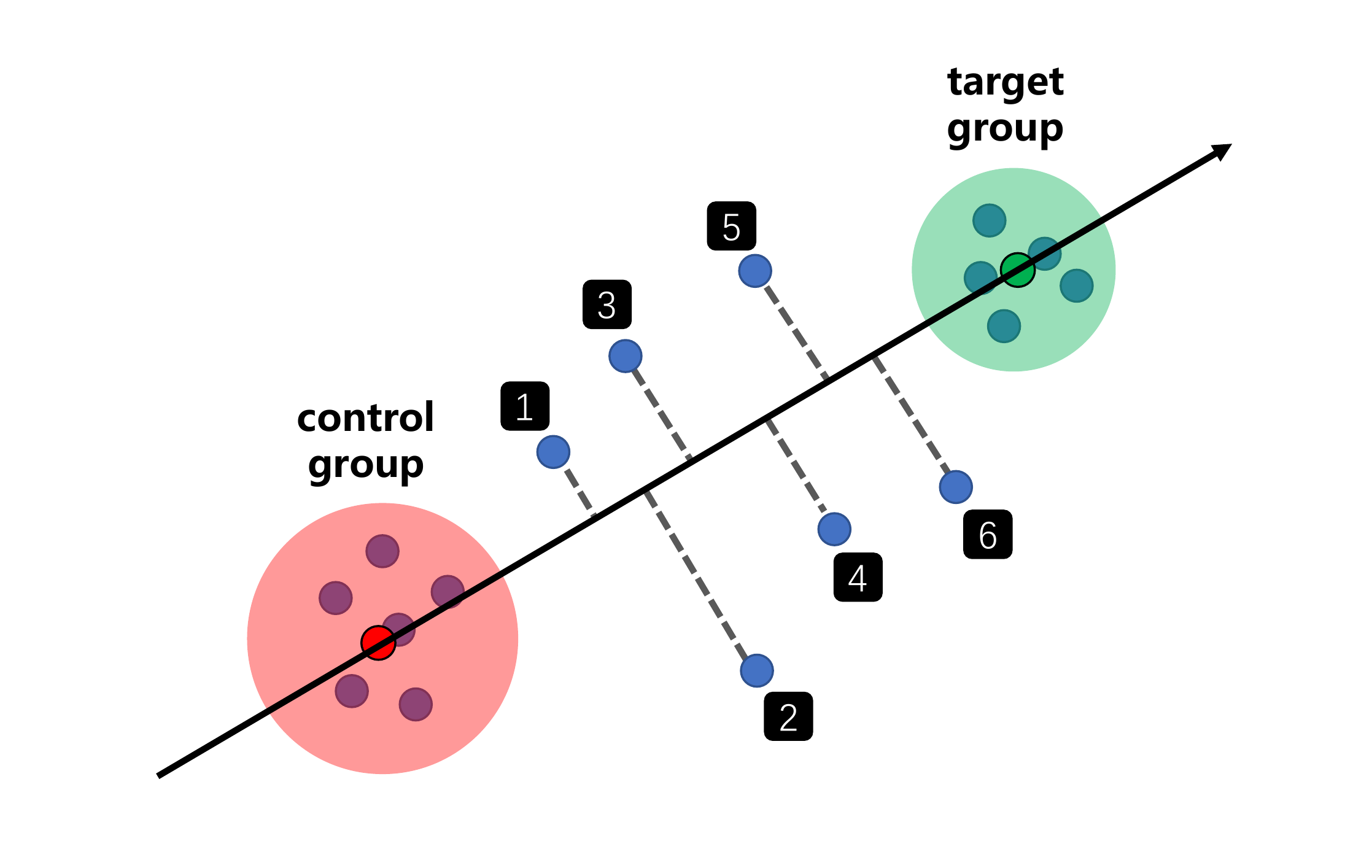}
  \caption{Illustration for SemanticAxis. SemanticAxis is a vector that passes through the center of two groups of high-dimensional data selected by the analyst. It serves as a metric for the semantic strength of data points whose rankings according to their projected position on the axis.}
  \label{fig:axis_construction}
\end{figure}

On the one hand, based on the idea of contrastive analysis, SemanticAxis can explain the semantics of an arbitrary region, such as a cluster or some outliers, by simply selecting it as the target group and meanwhile selecting another region, such as the majority of the remaining points or another cluster, as the control group. The differences on each dimension between the two group of points highlight the semantics of the target group (corresponding to the first task); on the other hand, the semantics of the axis changes in concept or strength along its direction (\autoref{fig:axis_construction}), e.g., the semantics may transfer from ``male" to ``female" for words, or transfer from ``excellent in natural sciences" to ``advanced in humanities" for students. Hence, SemanticAxis can be viwed as a metric in which data points are sorted by their projected position on the axis (corresponding to the second task).

In addition, we designed and implemented a visual analysis system (\autoref{fig:interface}), in which several visual components and interactions enhance the adaptability of SemanticAxis facing complex real-world scenarios. For example, weight editor is used for attribute refinement in multi-attribute rankings and reduced space re-construction; ranking rows supports combined filtering and detail inspection, making it possible for analysts to further validate the insights gained through the SemanticAxis.

We summarize our contributions as follows:
\begin{itemize}
  \item We proposed SemanticAxis, a technique that merges the task of feature understanding and weighted ranking of multi-attribute data into a united exploration context;
  \item We designed a visual analysis system, in which easy-to-use visual components and concise interactions accommodate the SemanticAxis to  complex analysis scenarios;
  \item We presented some interesting and valuable findings on academic strength distribution in computer science, such as the differences between institutions at different levels are markedly different.
\end{itemize}

\section{Related Work}
\subsection{Semantic axis techniques}
We refer to the technique of using vectors and their linear expressions to encode semantics and their transitions as semantic axis technique. Analogies in word embedding~\cite{mikolov2013distributed}, such as man-woman, can be regarded as semantic axes. Heimerl et al.~\cite{heimerl2018interactive} helped understanding semantic differences between two corpora by checking word distribution in a cartesian coordinate system that is spanned by two semantic axes which is trained from these two corpora and describe the same concept. Liu et al.~\cite{liu2017visual} expanded the descriptors of a defined semantics by detecting words that fall on each side of its corresponding word vector. Explainers~\cite{gleicher2013explainers} and InterAxis~\cite{kim2015interaxis} learned linear functions from analysts' continuous decisions to model the relationship between data attributes and abstract concepts in their minds. Our SemanticAxis follows a similar technical roadmap to the InterAxis. However, the tasks, hypotheses, and motivation of the two are completely different. To be specific, the InterAxis emphasizes on creating an interpretable reduced space by semantic modeling, but we focus on understanding the generated reduced spaces and its combination with data ranking.

\subsection{Understanding the DR results}
For image or text, it is a simple and effective way to interpret their DR results by directly~\cite{tenenbaum2000global} or interactively~\cite{heimerl2016docucompass, kim2016topiclens, li2019galex} drawing the original data as annotation in reduced space. For other types of high-dimensional data, Ji et al.~\cite{ji2019visual} applied parallel coordinates plots to identify hidden semantic features associated with recognized clusters. Cavallo et al.~\cite{cavallo2018visual} explained the characteristics of a local region by ploting semantic curves of various dimensions centered on a certain data point. However, gererally, their approach requires out-of-simple support of the DR algorithm they used. Axisketcher~\cite{Kwon2016axisketcher} allowed analysts to draw a discretionary curve in reduced space and then helped them understanding the meaning of region that the curve passes through by expressing it as a combination of multi-segment linear functions. Stahnke et al.~\cite{stahnke2015probing}, Liu et al.~\cite{liu2019latent}, and we all uncovered the feature of local areas by analyzing the differences between the points inside and outside the areas. Compared to the former, we assign the differences a semantic explanation and apply it to the entire data set for rankings. Compared to the latter, we offer multiple interactions towards exploring the constructed semantics in depth. For example, we design a brush filter to support examing the distribution of semantics in reduced space with different granularity.

\subsection{Interacting with DR model}
We highlight two kinds of interactions in the context of interacting with the DR model in reduced space: Parametric Interaction (PI) and Observation-Level Interaction (OLI) ~\cite{wenskovitch2017towards}. PI refers to manipulating parameters directly in order to create a new projection. This presents a difficulty to novice or non-mathematically-inclined analysts. Typical examples include slider bars from Andromeda (PI view)~\cite{self2016bridging}, Star Coordinates~\cite{kandogan2000star}, and SpinBox widgets from STREAMIT~\cite{alsakran2011streamit}. While OLI enables the analyst to directly manipulate the observations (data points), shielding the analysts from the complexity of the underlying mathematical models. Typical examples include StarSPIRE~\cite{bradel2014multi}, Paulovich et al.~\cite{paulovich2011piece}, and Mamani et al.~\cite{mamani2013user}. Recently, Jessica et al.~\cite{self2018observation} determined the differences, advantages, and drawbacks of PI and OLI, and drew the conclusion that these two serve different, but complementary. Semantic interaction is a similar concept to OLI whose interaction objects are also data points. It is just that the former puts more emphasis on the semantic interpretation of interaction. Semantic interaction follows the human-in-the loop pipeline ~\cite{endert2014human}, in which analysts spatially interact with data models directly within the visual metaphor using expressive interactions, and then system interface provides visual feedback of the updated model and learned parameters also within the visual metaphor~\cite{endert2012semantic}. For example, Endert et al.~\cite{endert2012semantic, endert2011observation} enabled analysts to interactively generate a well-interpreted document space by semantic interactions, such as moving, highlighting, grouping, and annotating documents. These interactions are interpreted and mapped to the underlying parameters of a force-directed model~\cite{endert2012semantic} or a weighted MDS model~\cite{endert2011observation}. In our SemanticAxis system, analysts can rebuild projection by directly modifying attribute weights, which is clearly a PI. The interaction of creating a semantic axis by selecting two groups of points belongs to neither PI nor OLI, since it does not interact with the DR model.


\subsection{Multi-attribute rankings}
Gratzl et al.~\cite{gratzl2013lineup} summarized some common visual designs for multi-attribute rankings, including spreadsheet~\cite{fewdesigning}, point-based, line-based (e.g., parallel coordinates plot~\cite{inselberg1985plane}, slope graph~\cite[p.156]{tufte2001visual}, and bump chart~\cite[p.110]{tufte1990envisioning}), and region-based (e.g., table with embedded bars~\cite{rao1994table}, multi-bar chart, and stacked bar~\cite{few2012show}) techniques. We adopted a line-based tenique: parallel coordinates plot, because it supports comparing the rankings of the same data point among various dimensions, multiple ranking criteria, and different time periods. Besides, dynamic weight adjustment is a widely used attribute refinement method~\cite{gratzl2013lineup, weng2018srvis, carenini2004valuecharts, wall2017podium}. We implemented it in our system to assit analysts in customizing desired ranking criteria.

\section{Methodology}
As introduced in the Introduction section, in order to understand the semantics of a target region in a reduced space, the analyst needs to construct a target group and a control group. We denote the vector of the i-th and j-th data point in the constructed target group and control group as $\boldsymbol p_i^t,\;\boldsymbol p_j^c\in R^{1\times N}$, respectively, and denote a weight vector that describes the importance of each attribute as $\boldsymbol\omega\in R^{1\times N}$ (sum up to 1). N represents the number of attributes. As shown in \autoref{fig:axis_construction}, we define the SemanticAxis $\boldsymbol v\in R^{1\times N}$ as the weighted vector that passes through the center of these two groups of points:
\[ \boldsymbol v\;=\;(\;\frac1T\sum_{i=1}^T\boldsymbol p_i^t\;-\;\frac1C\sum_{j=1}^C\boldsymbol p_j^c\;)\;\odot\;\boldsymbol\omega \]
where $\odot$, L, and R represent the Hadamard product (also known as the element-wise product) and the number of points in the two groups, respectively. Therefore, in essence, our SemanticAxis is a linear combination of the original attributes.

In high-dimensional space, as we walk along an attribute axis, the semantic strength it expressed increases or decreases monotonically. This is also true for the axis formed by a linear combination of attributes, as long as the linear combination indeed conveys an interpretable semantics affirmed by analysts. For example, consider a multi-arrtibute data that records the academic performance of institutions (data points) in areas (attributes) of computer science and an axis $\boldsymbol v$ is a linear combination of all AI-related areas, then the semantics of this AI-axis can be interpreted as the comprehensive performance of institutions in AI. For an arbitrary institution $\boldsymbol p$ in high-dimensional space, its projection position $proj_{\boldsymbol v}^{\boldsymbol p}=\frac{\boldsymbol p\boldsymbol\cdot\boldsymbol v}{\Arrowvert\mathbf v\Arrowvert}$ on the axis $\boldsymbol v$ can be used to measure its performance. In this way, SemanticAxis can be regarded as a ranking criterion based on its represented abstract semantics, in which data points are sorted by their relative projection positions on the axis. It is exactly what the multi-attribute rankings required. 



SemanticAxis utilizes the idea of contrastive analysis to interpret the semantics of any part of the reduced space. Attributes that have significant numerical differences between the target group and control group describe the characteristics (semantics) of the target cluster. Depending on the selection of the control group, semantic axis can be divided into two types: 
\begin{itemize}
  \item \textbf{Unipolar semantic axis}, whose control group consists of the majority of the remaining data points (no need to be exactly precise). In this case, the semantic axis only takes a single semantics and its semantic strength is simply getting stronger or weaker along the axis. Take the above AI-axis as an example, institutions on the left/right always perform better in AI than those on their right/left. It is worth noting that being strong in semantic strength (or to say holding good semantic performance) does not mean being evenly strong in all involved attributes. On the contrary, it may only be strong in partial attributes.
  \item \textbf{Bipolar semantic axis}, whose control group is another cluster. In this case, each end acts as a control group for the other end. As a consequence, each end holds a unique semantics that represents the characteristics of the corresponding cluster. The semantics gradually transfers along the axis. It is worth noting that, in this case, the projection position of data points reflect their \textbf{relative} instead of absolute differences on semantic performance between the two ends. Therefore, nodes toward one end are those whose semantic performance at the current end is far stronger than that at the other end, and nodes located at the middle of the axis are those whose semantic performance at both ends are similar. Imagine a ``AI - Theory" axis, whose two ends consist of institutions that perform excellent in AI and Theory, respectively, then the semantics of the AI-end is ``performance in AI is much better than that in Theory", while the Theory-end is the opposite. For those institutions located in the middle, their performances in AI and Theory are similar, either both excellent or both weak.
\end{itemize}


\begin{figure}
  \centering
  \includegraphics[scale=0.36]{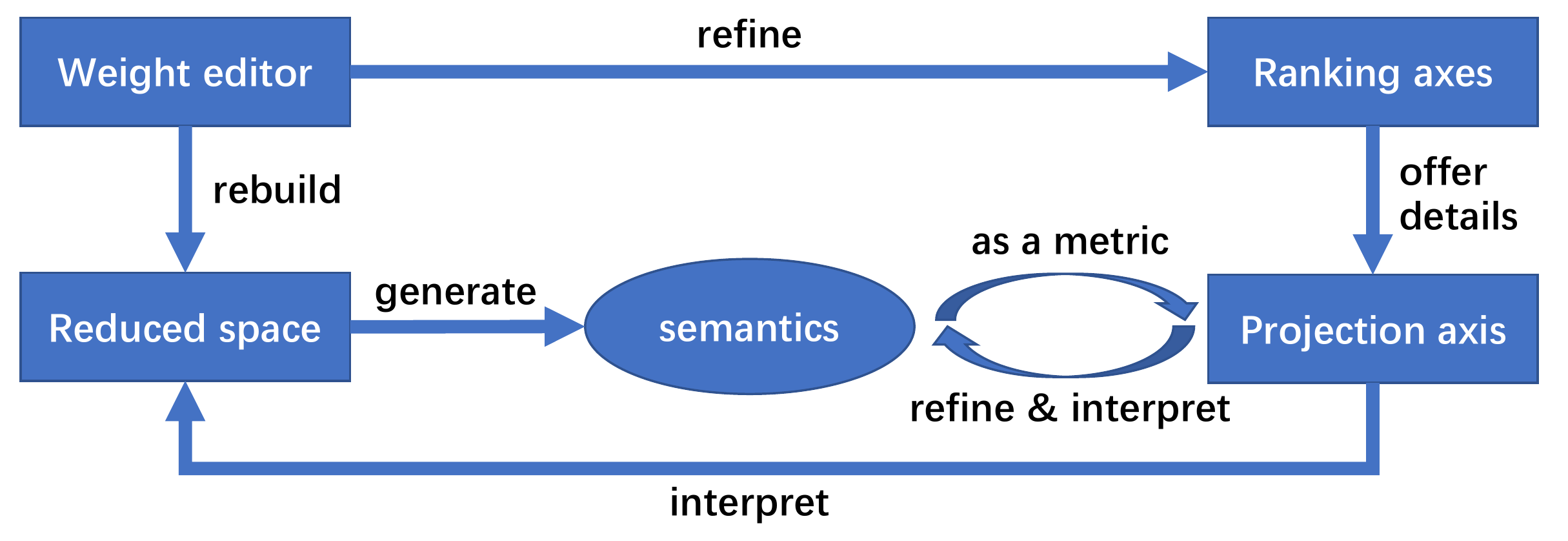}
  \caption{Relationships between four main components of our system.}
  \label{fig:framework}
\end{figure}

\section{Interface and interactions}
To enhance the capabilities and applicability of our SemanticAxis, we designed and implemented a visual analysis system. In this section, we introduce the visual design and interaction design of each component and the cooperations between them. The relationships between the four main components are summarized in \autoref{fig:framework}. 

\paragraph{Reduced space}
Reduced space uncovers the distribution of weighted high dimensional data through embedding them into a two-dimensional plane. The DR algorithm we used in our paper is t-SNE~\cite{maaten2008visualizing}, since it can give a visually explicit DR result where the potential clusters are usually well separated without losing the details of individual cluster. In fact, any DR algorithm can be used here, we will talk about this in detail in the Discussion section. In reduced space, each circle represents a data point, its radius is proportional to its weighted score given by $\sum_{i=1}^N\boldsymbol p \cdot \boldsymbol\omega$, where $\boldsymbol p$ and $\boldsymbol\omega$ are the data point vector and the weight vector, respectively. Analysts are free to lasso two sets of points in the reduced space as two ends of a semantic axis. The DR algorithm and the lasso interaction ensure the selected data group usually hold a stable and meaningful semantics, which is a precondition for the created semantic axis to be interpretable.

\paragraph{Weight editor}
Analysts can adjust the weight of each attribute in weight editor by changing the length of its corresponding rectangle. With the constraint that the sum of the weights remains 1, as an analyst increases (decreases) the weight of one attribute, the others decrease (increase) equally until one drops to zero. Compared to the design in which each attribute can be adjusted independently without the constraint and the linkage, our design greatly reduces the amount of operation needed for ignoring numerous undesired attributes. Weights adjustment is necessary for the following purposes:
\begin{itemize}
  \item First, to avoid the situation where an analyst is intent to describe a concept in mind but has difficulty finding its embodiment in the initial reduced space. For example, an analyst wants to construct a semantic axis that indicates how strong an institution is in computer theory. A conventional idea is to select institutions that excel in computer theory as one end and the rest as another end. However, the target institutions may be scattered throughout the current reduced space which prevents the analyst lassoing them. At this point, the analyst can change the weight of relevant attributes to reshape the reduced space so that the target institutions are clustered together.
  \item Second, to implement attribute refinement of multi-attribute rankings. On the one hand, analysts can build ranking criteria according to their own preferences, such as increasing the weight of the attributes that they care about; on the other hand, analysts can perceive the influence of a focused attribute on rankings by observing how the rankings change after adjusting their weights. The weighted ranking results are shown in weighted ranking row and we will mention it later.
\end{itemize}

\paragraph{Projection axis}
In projection axis, circles denote data points, radiuses are proportional to their weighted scores, x-positions are scaled according to their projection positions on the current semantic axis. A force-directed algorithm is utilized to prevent the overlap of data points. Rectangles represent all attributes of data, we call them attribute rectangles. Their height is proportional to the absolute value of its corresponding element of the current semantic axis vector $\boldsymbol v$. They are evenly placed and sorted by the absolute value and the sign of the value they bind to, which means the rectangles with large absolute value are listed near the ends and the rectangles with positive/negative value are placed above/below the axis. Attribute rectangles are used to illuminate the semantics of the current axis. For example, in Figure 1, the two rectangles at two ends reveal that the current axis describes the performance difference of institutions in visualization area and computer graphics area. Notice that $\boldsymbol v$ is scaled by the weight vector $\boldsymbol\omega$ element-wisely, as shown in the Methodology section. So there are two possible situations where several attribute rectangles are too short to be visible: 1. the two selected data groups of the current axis show no difference on these attributes; 2. the weight of these attributes are set pretty low by analysts in the weight editor. 

Analysts can fine-tune the semantics by slightly changing the height of rectangles. The benefits of this interaction are two-fold: first, it allows analysts to eliminate the deviation (values of the irrelevant non-zero attributes) between the constructed semantics and the expected semantics; second, post-adjustment makes it unnecessary for analysts to painstakingly picking the data points during constructing axis in the reduced space, which improves the efficiency of analysis. Projection axis wii be updated automatically each time finishing lassoing two groups of points in reduced space.
    
\begin{figure}
  \centering
  \includegraphics[scale=0.27]{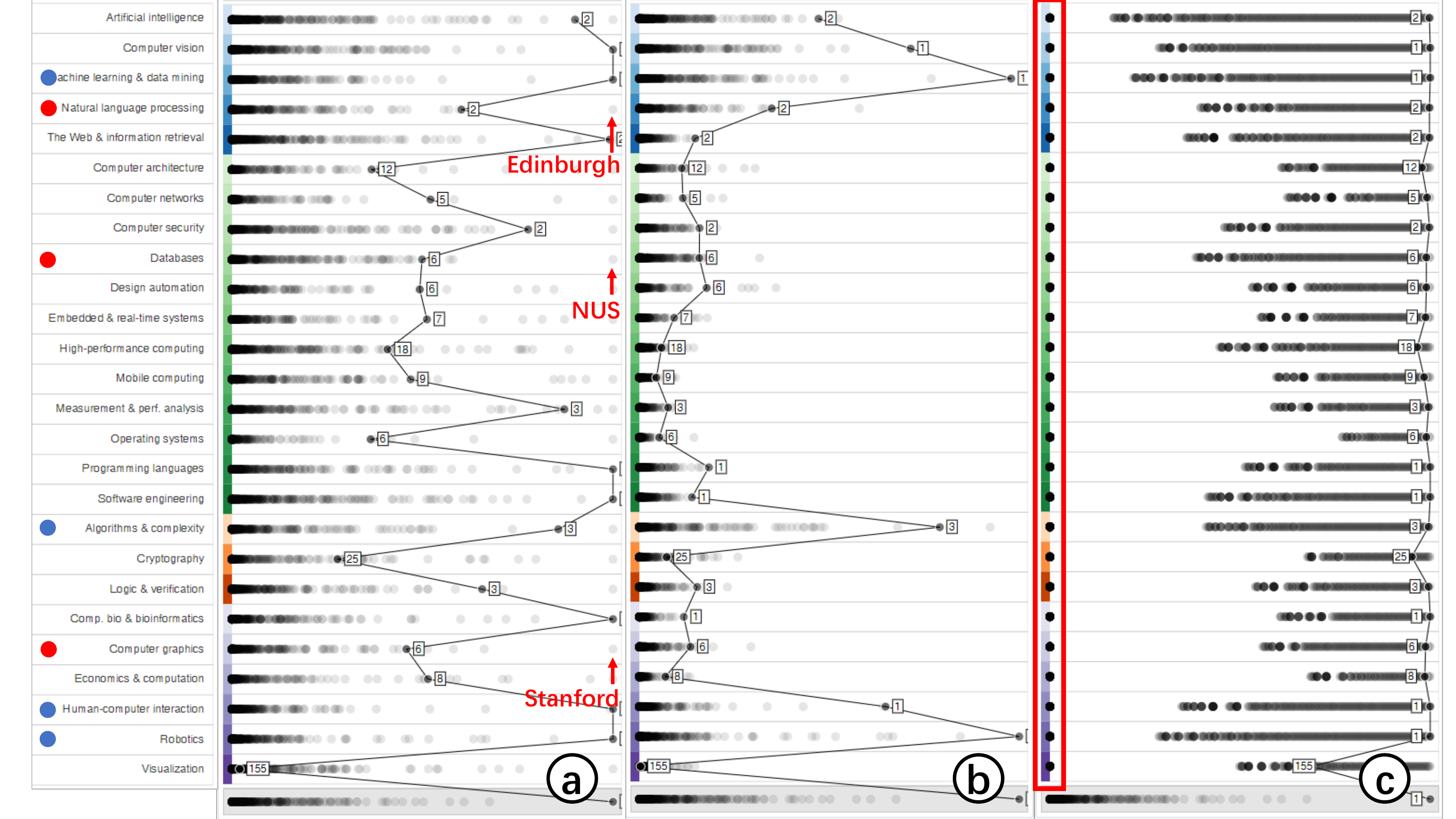}
  \caption{Illustration of the three scale strategies in attribute rows: scale with values of individual attribute, values of all attributes, and rankings. {\normalsize{\textcircled{\small{a}}}} demonstrates that, for most areas of computer science (especially for the areas labeled with red node), there are a handful of institutions that perform noticeably well. {\normalsize{\textcircled{\small{b}}}} shows that, the scale of the top conferences of blue-node areas is much larger than that of other areas. The large blank on the left of {\normalsize{\textcircled{\small{c}}}} indicates that, for all areas, there are a significant number of institutions (in the red box) failed to publish in the corresponding top conferences.}
  \label{fig:ranking_rows}
\end{figure}

\begin{figure}
  \includegraphics[scale=0.30]{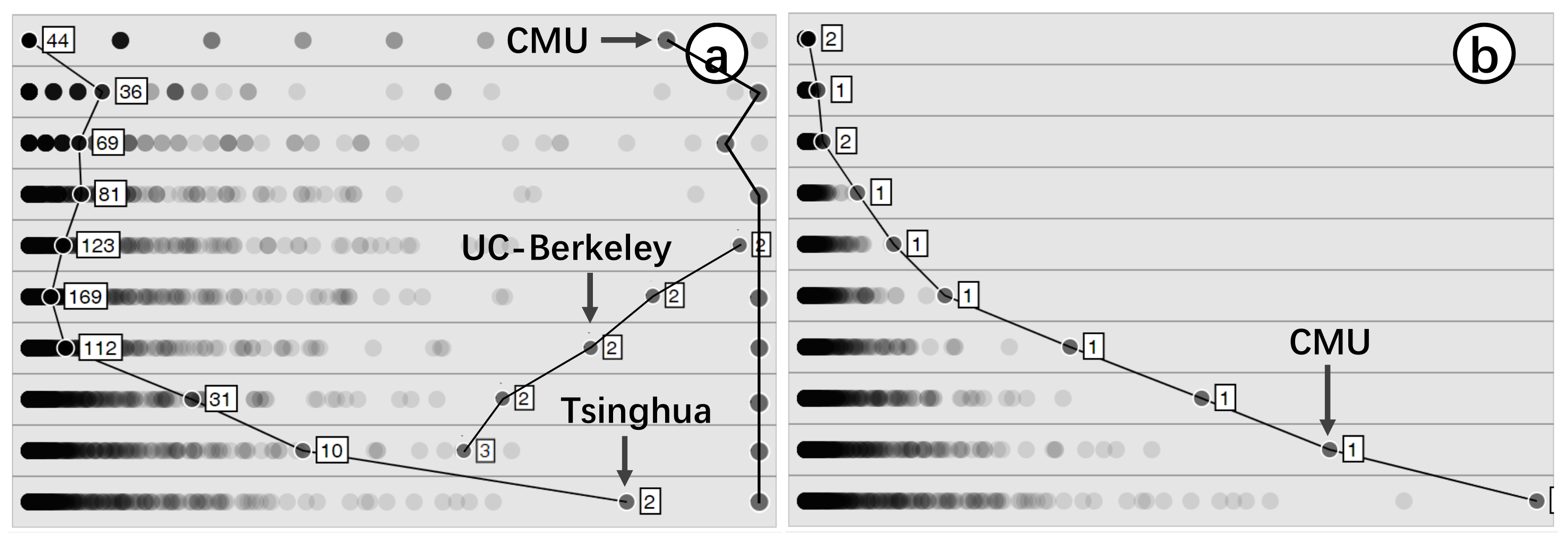}
  \caption{Example of a weighted ranking row unfolded by time stages. We split the period 1970 to 2020 into 10 segments with a 5-year time interval. Each row reflects the comprehensive strength of institutions in the corresponding time period. Same as in the attribute rows, analysts can select local scale ({\normalsize{\textcircled{\small{a}}}}) or global scale ({\normalsize{\textcircled{\small{b}}}}). {\normalsize{\textcircled{\small{a}}}} indicates that the gap between the first (basically, Carnegie Mellon University (CMU)) and the second (basically, UC-Berkeley) institutions gradually widen since 1990, until Tsinghua University rushed to the second. {\normalsize{\textcircled{\small{b}}}} presents the rapid development of computer science in the past decades.}
  \label{fig:filter_results_row}
\end{figure}

\paragraph{Ranking rows}
Ranking rows component has two functions: one for validating observations obtained from projection axis by further checking details of data points and the other for supporting several multi-attribute ranking tasks with simple designs. From top to bottom, all rows are divided into three groups: filtering results row, attribute rows and weighted ranking row. Attribute rows are similar to parallel coordinates plot where the position of a data item in each row indicates its performance in the corresponding attribute. The space for those emphasised attributes, like the visualization and computer graphics in \autoref{fig:interface}, is expanded, while all data points are spread out, enabling individual checking. We design three linear scales to compute the position, and each of them focuses on an aspect of the data. The first scale is a local scale whose domain is the extent of attribute values of the current attribute. It is suitable for comparing the distribution of attribute values between attributes (see \autoref{fig:ranking_rows} (a)). The second scale is a global scale whose domain is the extent of attribute values of all attributes. It is suitable for comparing the extent of attribute values among attributes (see \autoref{fig:ranking_rows} (b)). The third scale is a local scale. It places data points by their rankings in current attribute. It can reveal the distribution of the rankings (see \autoref{fig:ranking_rows} (c)). The polyline crossing all attribute rows connects the same data item and uncovers the characteristics of the data item. In each row of attribute rows, a filter can be created by brushing. Created filters are combined by the ``and'' operator, and the final filtering results are presented in the filtering row. Weighted ranking row is used to present the weighted ranking results of data points based on their weighted scores. Analysts can split the row into time slices to track changes in the value or ranking of data items in different stages (see \autoref{fig:filter_results_row}).

\begin{figure}
  \centering
  \includegraphics[scale=0.30]{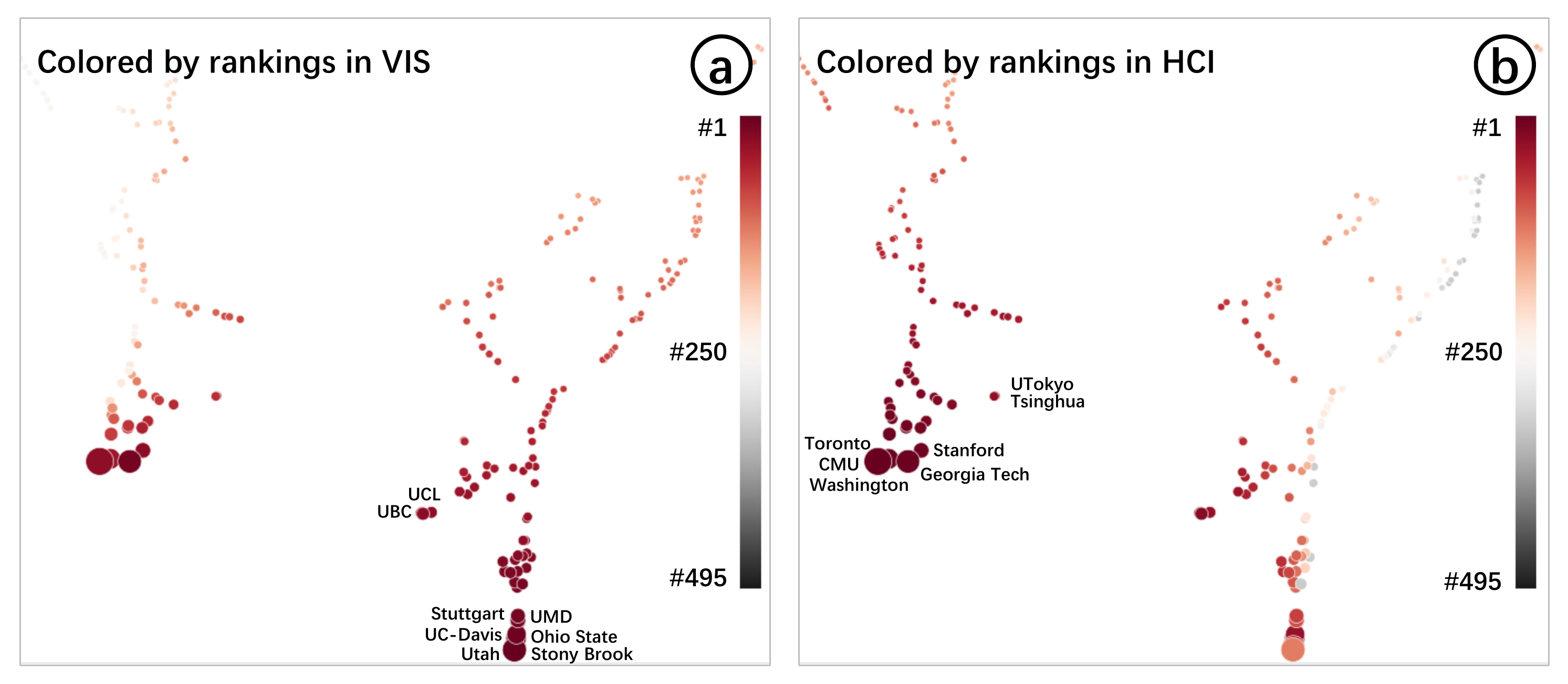}
  \caption{Examples of coloring items by their values or rankings on selected weighted attributes. In {\normalsize{\textcircled{\small{a}}}} and {\normalsize{\textcircled{\small{b}}}}, institutions (points) are colored according to their rankings in VIS area and HCI area, respectively. We can see that the bands on the left and right roughly represent institutions that perform well in HCI and VIS, respectively.}
  \label{fig:coloring}
\end{figure}

\begin{figure}
  \centering
  \includegraphics[scale=0.47]{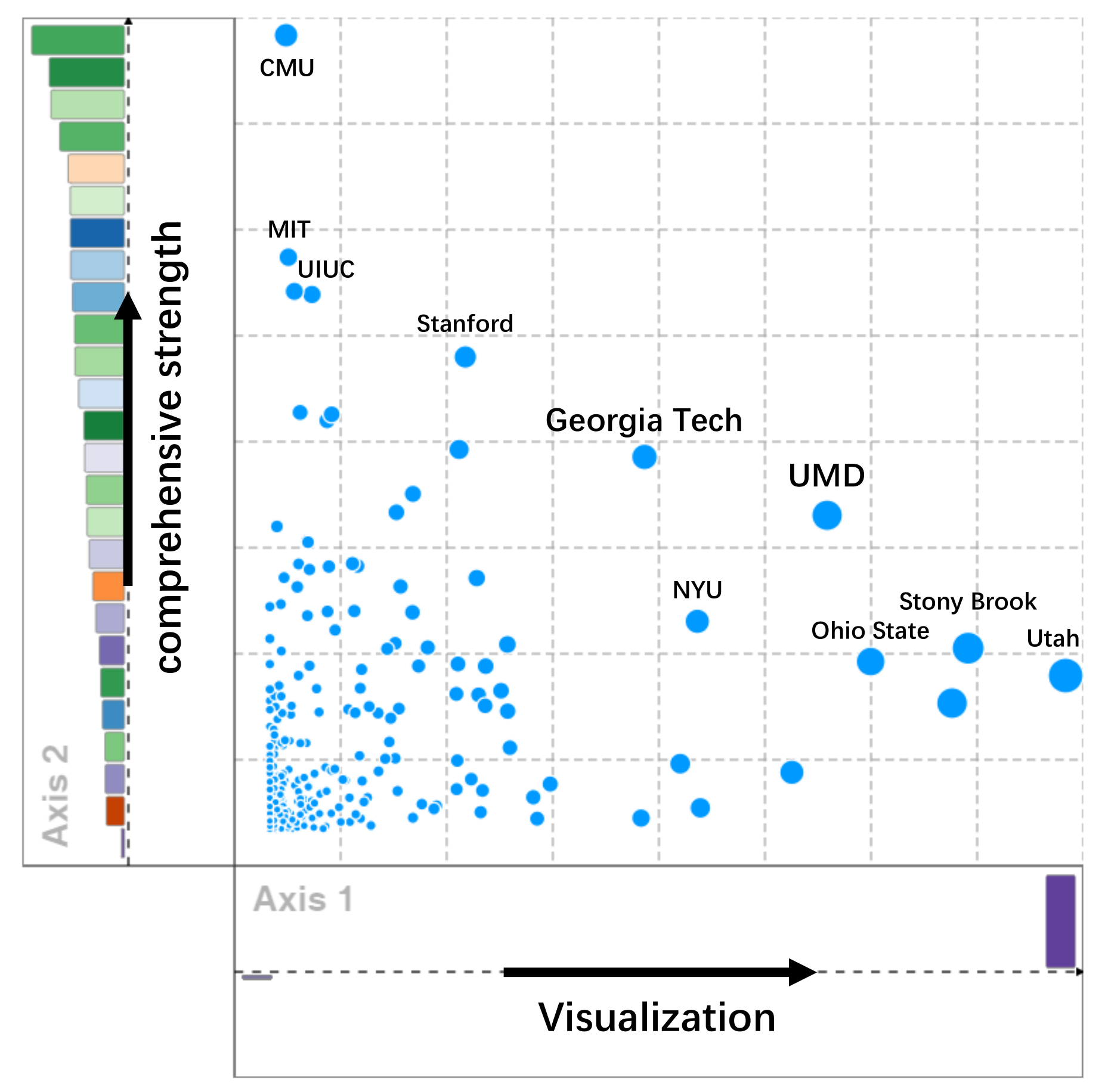}
  \caption{A simple example of a composite custom semantic space. The distribution of institutions (points) indicates that the correlation between the performance in the VIS and the comprehensive strength is small. Georgia Tech and College Park are among the few options for institutions that have achieved great rankings in both aspects.}
  \label{fig:two_axes}
\end{figure}

\paragraph{Interactions between components}
After adjusting the weights in weight editor, analysts can click the update button to update reduced space and ranking rows. To examine the distribution of single or composite attributes in reduced space, analysts can check the corresponding attributes in weight editor, and accordingly, the circles in the reduced space are colored by their weighted scores (the case of composite attributes) or attribute values (the case of single attribute) on the checked attributes (see \autoref{fig:coloring}). 
To get rid of the limitation that analysts can only learn the data distribution in a single semantic axis, we designed a two-dimensional composite semantic space whose x-axis and y-axis are two created semantic axes, respectively. Its creation process is stated as follows: during the exploration, analysts are allowed to save the created semantic axis and check it by hovering its corresponding filled square at the top right corner of reduced space. Once two axes are saved, analysts can click the right-most icon to show the custom composite semantic space in a pop-up window. The x and y coordinates of data points inside the space correspond to their projection positions on the two saved axes. For example, as shown in \autoref{fig:two_axes}, analysts can identify institutions that excel in VIS area while having strong comprehensive strength. In projection axis, we allow analysts to brush some data points, and the selected points will be highlighted in reduced space and attribute rows. The observed granularity can be adjusted according to the width of brush. This interaction is simple but very useful, giving analysts the ability to scrutinize the distribution of captured semantics in reduced space. We will elaborate on this in Case Study.

We notice that it is hard for analysts to remember the semantics of all clusters they have ever explored. They may repeatedly construct similar semantic axes to examine the same cluster, which greatly reduces the efficiency, especially when there are numerous clusters and their boundaries are not clear. In order to alleviate the memory burden of analysts and prevent repetitive operations, we designed a storage mechanism called checkpoints. It allows analysts to save all information of the new created semantic axis, including a snapshot of the whole projection axis, the lassoed data points, and the lassoed regions. Meanwhile, a circle representing the checkpoint is pinned at the center of each lassoed region. Analysts can choose to hide the checkpoints representing control groups. When hovering over a checkpoint, its saved information emerges; if the checkpoint is clicked, the information would be embedded back in appropriate panels for further viewing. In addition to solidifying knowledge during exploration, checkpoints can serve as landmarks for reduced space, providing focus points and aiding to navigations~\cite[p.156]{ware2013information}, for example, guiding analysts to check unexplored regions~\cite{han2019visual}.

\section{Case Study}

\begin{figure}
  \centering
  \includegraphics[scale=0.26]{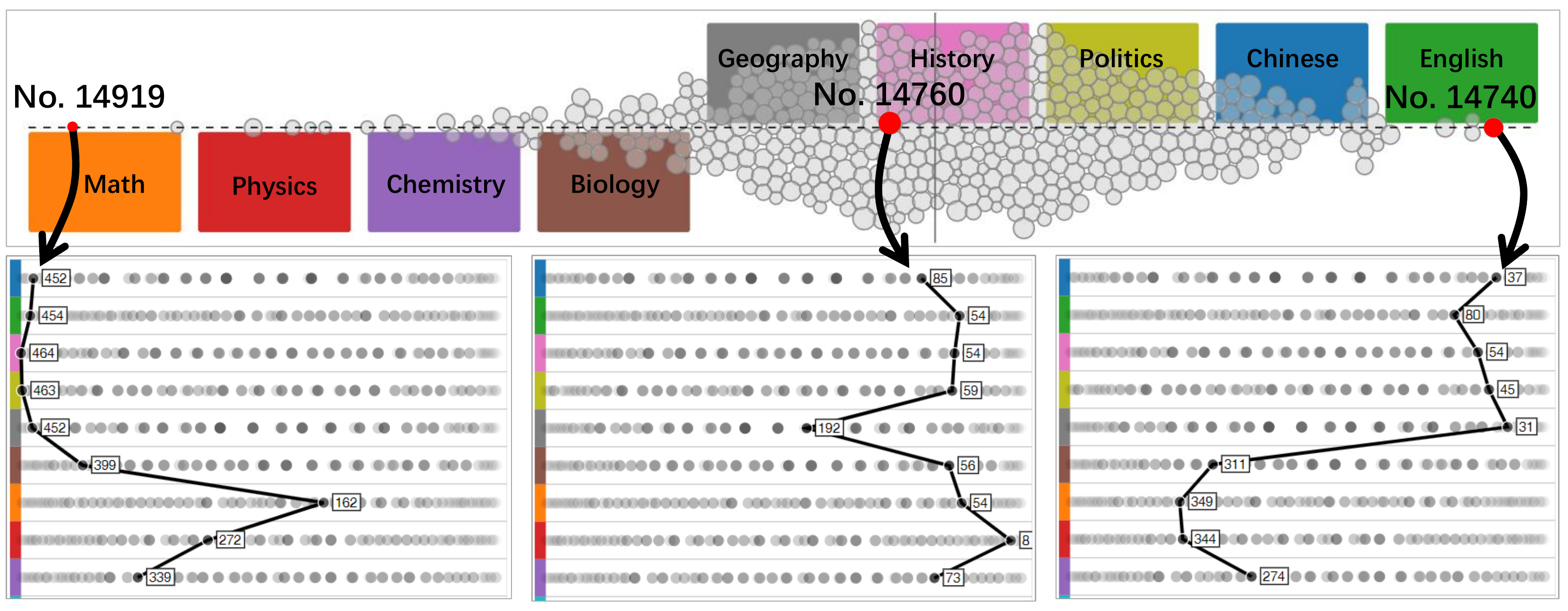}
  \caption{Constructed SemanticAxis for detecting biased students. Students located at two ends of the axis hold a serious bias. It is verified by the ranking details offered by the ranking rows.}
  \label{fig:case1_1}
\end{figure}

\begin{figure}
  \centering
  \includegraphics[scale=0.285]{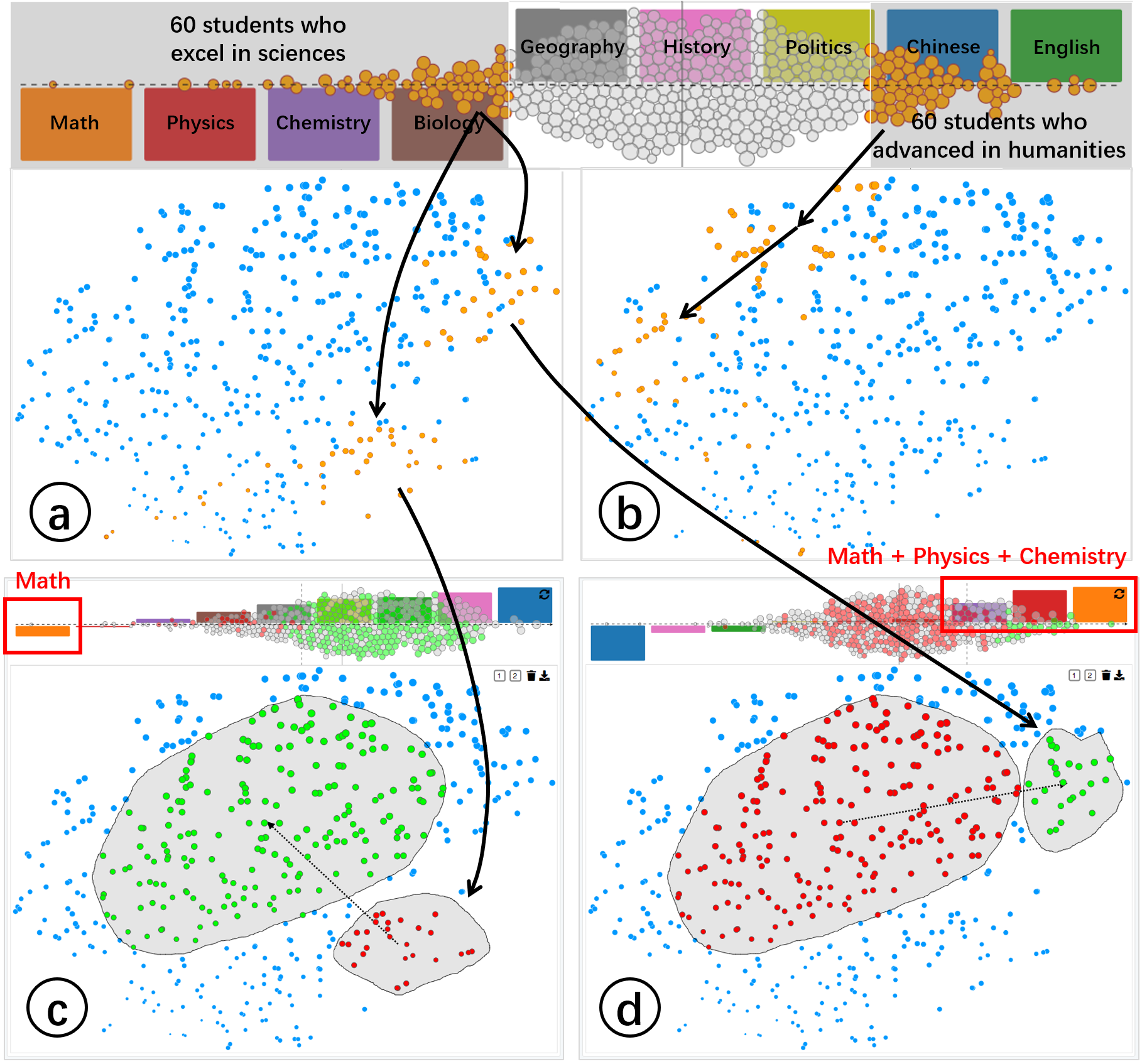}
  \caption{Biased students found by SemanticAxis are distributed on the lower right ({\normalsize{\textcircled{\small{a}}}}, partial to the sciences) and upper left ({\normalsize{\textcircled{\small{b}}}}, partial to humanities) sides of the reduced space. The differences between the two groups of students in {\normalsize{\textcircled{\small{a}}}} lay in that the former only excel in math ({\normalsize{\textcircled{\small{c}}}}) while the latter is good at math, physics, and chemistry ({\normalsize{\textcircled{\small{d}}}}).}
  \label{fig:case1_2}
\end{figure}

\subsection{Case 1: Score data of students}
In this case, we aim at finding biased students according to their scores in an exam. The biased students refer to the students who go overboard on partial subjects but perform weakly in others. For example, some students may excel in natural sciences but do badly in humanities. It is necessary for teachers to find out these biased students, because they should be given guidance towards balanced development or be encouraged to dive into their specialties. Our data records the scores of 494 students at a middle school in Ningbo in a final exam involving nine subjects. The original scores are converted to z-scores to ignore differences in the data distribution between subjects. In this case, we pay attention to the performance differences between natural sciences and humanities. Traditionally, we consider natural sciences to include mathematics, physics, chemistry, biology, and humanities to include Chinese, English, history, politics, and geography.

We expect to construct a semantic axis to measure the performance differences. A reasonable design would be to set one end means ``excel in the sciences" and the other represents ``do well in humanities". Since the target semantics is clear, we can construct the semantic axis in projection axis by dragging attribute rectangles directly. As shown in \autoref{fig:case1_1}, we set the value of the semantic axis on subjects in the two fields to a negative value and a positive value, respectively. Currently, the projection position of students on this axis should convey their degree of bias, that is, students at two ends may have a serious bias. To verify this, we select three representative students and check their rankings on each subject in ranking rows (\autoref{fig:case1_1}). We find that the No.14919 student ranked significantly better in natural sciences than humanities, the No.14740 student did the opposite, and the No.14760 student located in the middle of the semantic axis showed no obvious difference between the two. This indicates the correctness of the axis we constructed.

Next, we want to check whether the current reduced space captures the ``biased" semantics. We select 60 students at each end with brush filter and examine their distributions in the reduced space  (\autoref{fig:case1_2}). We find that these students are clearly clustered on the lower right side ({\normalsize{\textcircled{\small{a}}}}) and the upper left side ({\normalsize{\textcircled{\small{b}}}}), which suggests a positive answer to the previous question. Then we notice that the students prefer the sciences are distributed mainly in two distinct regions ({\normalsize{\textcircled{\small{a}}}}). Hence, we build two semantic axes to understand their differences. As shown in {\normalsize{\textcircled{\small{c}}}} and {\normalsize{\textcircled{\small{d}}}}, one region includs students who have a specific advantage in math alone, while the other region includ students who perform well in all subjects in the sciences, except for biology. 

It is worth noting that there is no obvious cluster or other features (clues) in the original reduced space, which may cause analysts have no idea where to start the exploration. We overcome this by allowing analysts to build an initial impression of the reduced space by exploring the semantics in their minds first.


\subsection{Case 2: Academic performance data of institutions}
In this case, we explore the academic performance and rankings of the whole world institutions in computer science. Potential users include decision-makers who manage and plan the subject of computer science and students who look forward to choosing an ideal school. Our data comes from CSRankings~\cite{Emerycsrankings} and is updated to April 2020. CSRankings devides computer science into 4 categories and 26 sub-areas. It scores 495 institutions on all areas based on their papers published in corresponding top conferences. These conferences are carefully chosen by senior domain experts. We set an individual linear scale for each area to range all scores between 0 and 100. Then we assess a synthesis score for institutions by weighted arithmetic mean on areas. 

\begin{figure}
  \centering
  \includegraphics[scale=0.38]{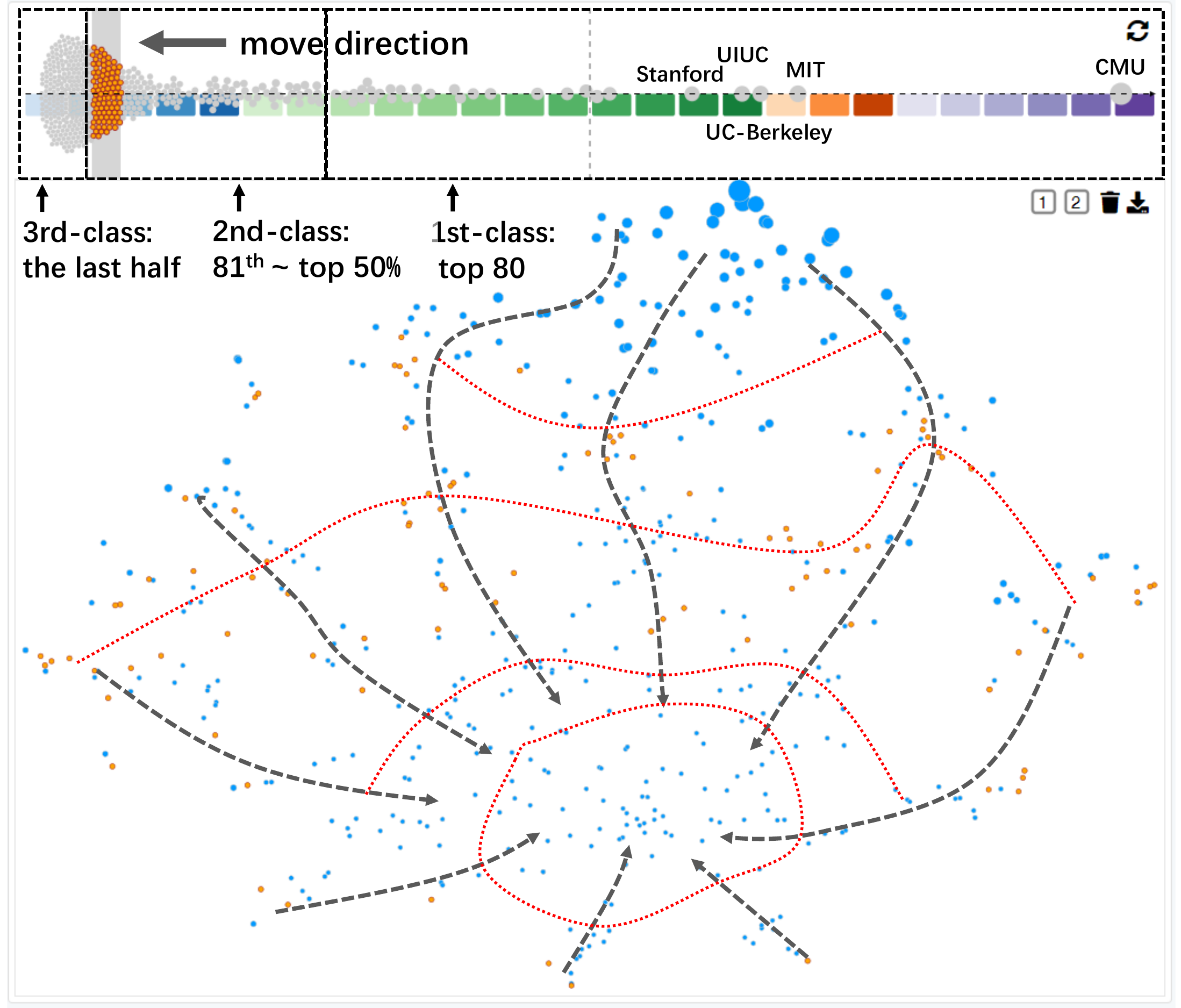}
  \caption{Illustration of the movement of selected institutions while moving brush filter from the right to the left. Black lines represent moving trajectories, and red lines signify arrival moments.}
  \label{fig:general_pattern}
\end{figure}

\begin{figure*}
  \centering
  \includegraphics[scale=0.35]{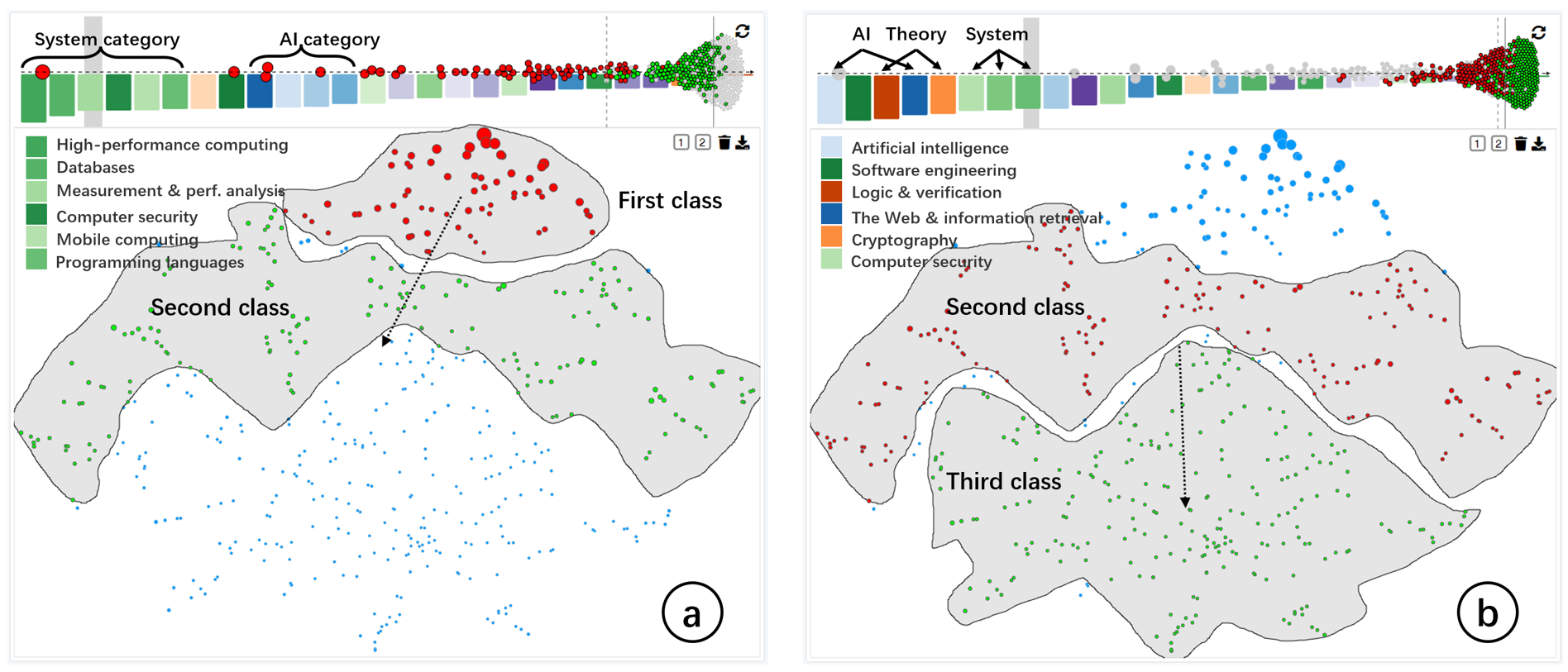}
  \caption{Institutions in upper class perform better than those in lower class in all areas. Nevertheless, the differences between the first (top 80) and second class (81th $\sim$ top 50\%) are significantly different from that between the second and third class (last 50\%). It is mainly reflected in the conspicuous forward movement of areas under AI and Theory category in the later.}
  \label{fig:class_comparison}
\end{figure*}

First, let us take a general perspective to check the distribution of institutions in reduced space. In the initial settings, the weight of all areas is 1/26, and the value of each attribute of initial semantic vector is equal. This means that currently, the position of points in projection axis and the size of the nodes in reduced space indicate the comprehensive strength of its corresponding institution. We discover a probable general pattern in the reduced space: the stronger the comprehensive strength of an institution, the higher its position. To verify this observation, we construct a narrow filter brush in the projection axis, and then move it from right to left by inches, while paying close attention to the movement of highlighted (brushed) points in the reduced space. We find, generally, the highlighted points move from up to down (\autoref{fig:general_pattern}), which means that the vertical direction of the initial reduced space almost represents the comprehensive strength. Hence, the general pattern is approximately true. 

Then we expect to understand the differences between different levels of institutions. As shown in the projection axis (\autoref{fig:general_pattern}), we divide all institutions into three classes: first (top 80), second (81th $\sim$ top 50\%), and third class (last half). Then we build two semantic axes with these three classes as endpoints (\autoref{fig:class_comparison}). Unsurprisingly, institutions in upper class perform better than those in lower class in all areas. Nevertheless, the main differences between the first and second class lay in areas under System category, while the main differences between the second and third class lay in areas under AI category and Theory category. 

We try to give a possible explanation for this finding. System category contains many old-line areas of computer science in which traditionally strong institutions have accumulated significant advantages. In contrast, most AI papers have been published over the past 10 years. The first class and second class almost stood on the same starting line. Besides, deep learning has swept through the whole academic, mobilizing the enthusiasm of almost all researchers in relevant fields, which makes the papers published in top conferences are no longer concentrated on a few top institutions (such as the first class institutions). The second-class institutions have participated in the competition extensively. However, the third class is still unable to catch the deep learning fast train, resulting in a widening gap in AI with the second class. As for areas in Theory category, their leading institutions are in the second class, leading to a comparable status between the first and second class while a relatively large gap between the second and third class.

\begin{figure}
  \centering
  \includegraphics[scale=0.4]{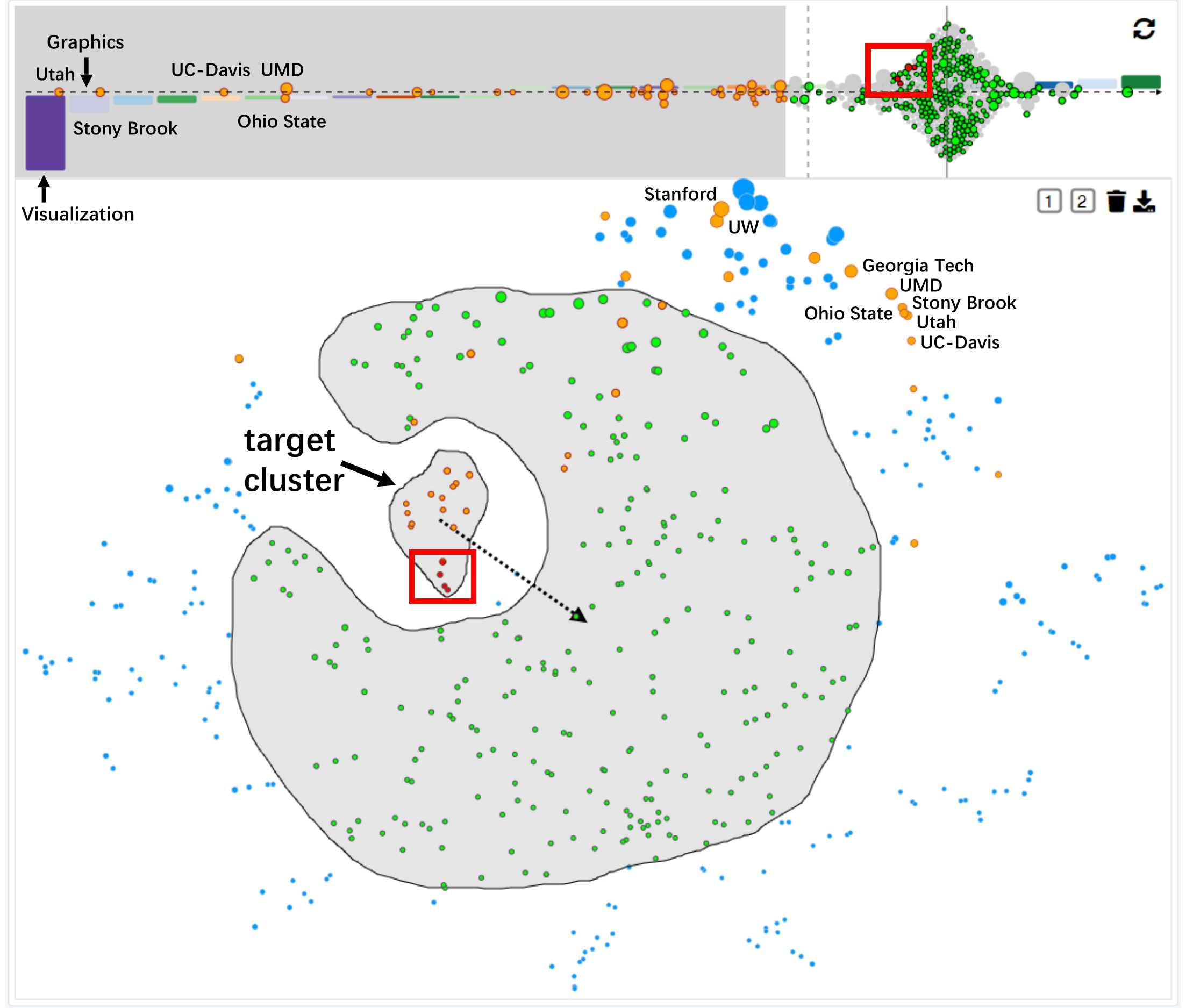}
  \caption{Insights uncovered by checking the lassoed cluster: four institutions (shown in the red box) should not have been selected; the exact semantics of the focused cluster is that it includes institutions that are nearly only excellent in the VIS area.}
  \label{fig:understand_cluster}
\end{figure}

Next we notice a cluster in the middle of the reduced space and intend to understand its meaning (\autoref{fig:understand_cluster}). We lasso this cluster and almost all the other points as the two ends of a SemanticAxis. In projection axis, we see that all the other areas are close to zero compared to the visualization area, which means that the feature of the cluster is ``excellent in visualization area" and institutions with the same feature should be placed near the left. Further, we notice that there are four institutions in the lasso but away from the left, which means the current lasso are not accurate and these four institutions should not have been included in the cluster. Conversely, several institutions, such as Maryland-College Park (UMD), Stony Brook, Utah, and UC-Davis, are not in the lasso, but at the upper right corner of the reduced space. We infer that these institutions are excellent enough to follow the general pattern. This suggests another underlying semantics of the institutions in the focused cluster: they are mediocre in most other areas. All of these inferences are verified with the details provided by ranking rows. Now, we have shown that our system can help analysts to understand the precise semantics of clusters, correct inaccurate cluster boundaries, and detect and interpret outliers.

\begin{figure}
  \centering
  \includegraphics[scale=0.43]{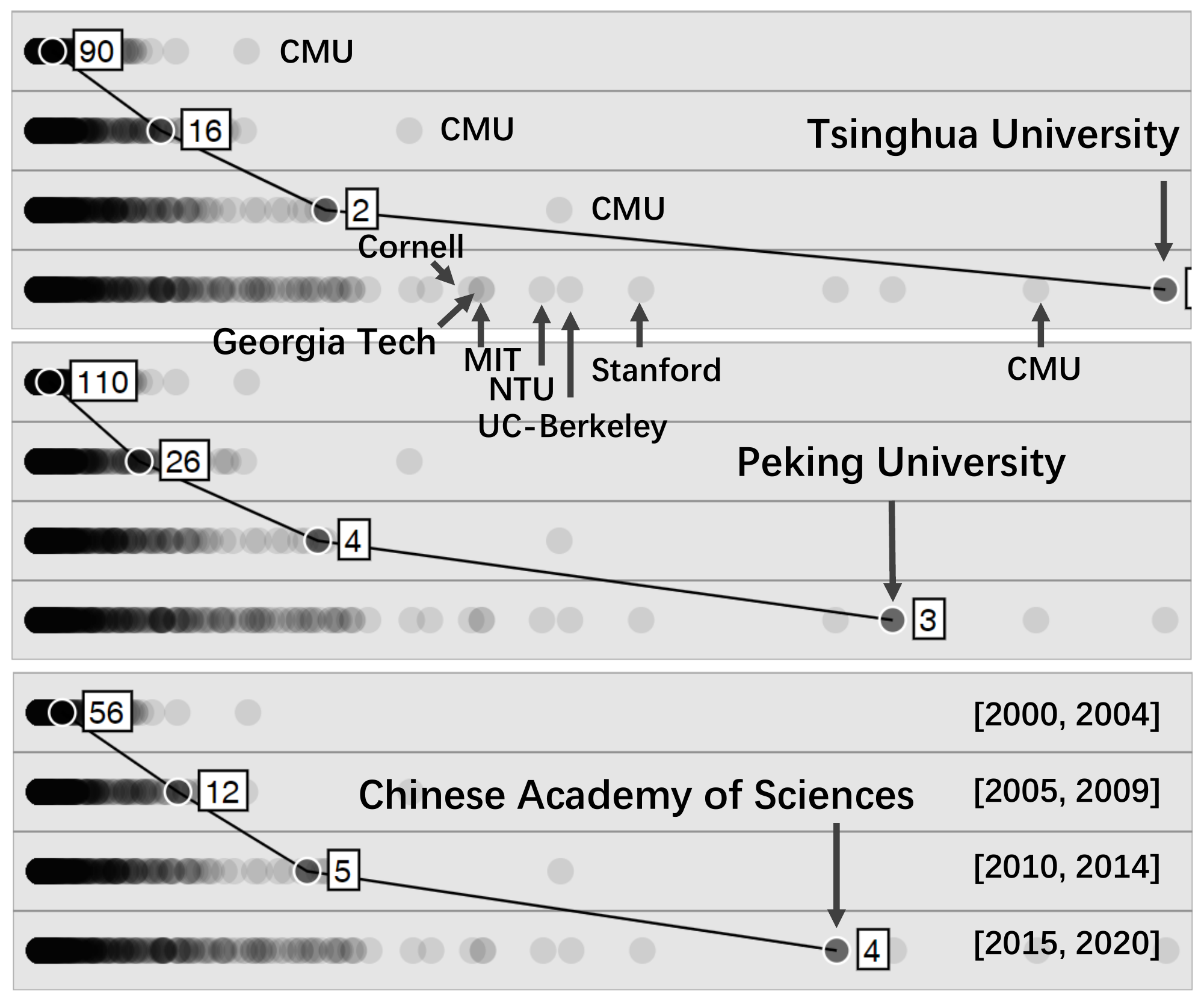}
  \caption{In the field of AI, several institutions in China have made great progress over the last 20 years. The competition between China and the US has become increasingly strong.}
  \label{fig:china_vs_us}
\end{figure}

Finally, we expect to check how the rankings have changed in the field of AI. We set the weight of each of the five areas under AI category at 20\% in weight editor and then divide the period from 2000 to 2020 into four segments at five-year intervals in weighted ranking row. We find that several Chinese institutions have made great progress (\autoref{fig:china_vs_us}). For example, Tsinghua University rose from 90th to 1st, Peking University rose from 110th to 3rd, and the Chinese Academy of Sciences rose from 56th to 4th. The top universities in the United States, such as CMU, Stanford, and UC-Berkeley, have always been among the best. The competition between China and the US in AI has become increasingly strong.

\section{Expert interviews}
In order to verify the effectiveness of our system in practice, we conducted an informal user study. We invited three front-line middle school teachers and three professors in charge of discipline construction at our college to take part in the analysis scenarios in Case 1 and Case 2, respectively. We spent 30 minutes explaining the goals, visual encodings, and interactions of our system and presenting the findings showed in the Case Study, then asked them to operate it for 30 minutes, and finally performed a 20 minute interview with each person. In the interview, they all agreed that our system is efficient, easy to use, and could help them get some valuable information that is difficult to grasp in their daily work. One teacher said, ``I know the general learning situation for most students in my class, but I can not tell the specific characteristics of each student. Your interactive visualization shows many clear and diverse results that I think I should review many times in my future work." ``... not only the individual level, the distribution shown in Projection Axis can reflect systematic biased among students", mentioned by another teacher. ``We rarely get such macro knowledge [referring to the systematic differences between different levels of institutions]" said one professor, ``It seems that focusing on hot areas is a feasible strategy for moving up the rankings quickly."

\section{Discussion and future work}

\paragraph{Non-linear semantics}
It is necessary to point out the caveats when analyzing nonlinear semantics using our linear SemanticAxis. We refer the semantics that analysts want to describe using the semantic axis the target semantics. It should be noted that only if the target semantics is linear, i.e., it can be described by a linear combination of the original attributes, the projection position of data points on our linear semantic axis is proportional to their strength on the target semantics. As the target semantics and the constructed semantic axes shown in \autoref{fig:case1_1} and \autoref{fig:general_pattern}, which are all linear. For nonlinear target semantics, such as the semantics implied in a cluster uncovered by nonlinear DR algorithms, the above proportional relation no longer holds. Constructed semantic axes even can not be used to roughly sort data points by their strength on the target semantics when the target semantics is highly non-linear. As in Case 2, the data points of the focused cluster are not at the left or right end of the constructed axis, i.e., the ranking function fails at this point. But this does not mean that our semantic axis technique is helpless against non-linear semantics. We can see that the semantics of the constructed axis ``performance in VIS area" is actually a linear approximation of the target semantics ``excellent in VIS area only". This suggests that our linear semantic axis can give significative hints (the cluster is related to VIS area) about the precise semantics of the target cluster. Besides, as introduced in Case 2, interactions of our semantic axis with other components allow the analyst to infer and confirm the exact semantics of the target cluster.

\paragraph{Choice of DR algorithm}
In fact, any DR algorithm can be applied in reduced space, its choice shares no relationship with the design of semanticAxis and the linearity of target semantics, as long as it can uncover the latent global and/or local structure of the high-dimensional data of interest. We chose t-SNE~\cite{maaten2008visualizing} because it tends to reveal more visually obvious local structures than other DR algorithms by emphasizing the different characteristics between latent clusters. For linear DR algorithms, like PCA~\cite{wold1987principal}, arbitrary direction in reduced space can be expressed as a linear combination of the original dimensions, which can be directly revealed by our semanticAxis; for non-linear DR algorithms, like t-SNE and UMAP~\cite{mcinnes2018umap}, our semanticAxis can provide a linear approximation of nonlinear semantics revealed by them, as we mentioned in the previous sub-section.

\paragraph{Scalability}
 One of the advantages of the linearity of our SemanticAxis is that the computation it involved is simple. Hence, SemanticAxis technique does not suffer from scalability problems. However, force-directed algorithm and dimensionality reduction algorithm, which are frequently used in our system, take on high complexity. For the former, it is an alternative to discretize the continuous position of data points by bins~\cite{rodrigues2017nonlinear}; for the latter, descending sampling and applying a more efficient dimensionality reduction algorithm (e.g. UMAP or PCA) are two mitigatory methods.

 \paragraph{Future work}
Future work involves three aspects: first, enhance the current SemanticAxis in analyzing non-linear semantics; second, promote the efficiency in checking the semantics of clusters by breaking the limitation that the current semantic axis only allows analysts to inspect clusters one by one (unipolar semantic axis) or two by two (bipolar semantic axis); third, embed our SemanticAxis into the human-in-the-loop analysis process, helping analysts to understand the model and add prior knowledge to the model in a positive feedback loop ~\cite{endert2011observation}.

\section{Conclusion}
In this paper, we propose SemanticAxis, a technique towards exploratory analysis of multi-attribute (or multivariate) data, and then we present a visual analysis system with the SemanticAxis at its core. SemanticAxis characterizes abstract semantics by a linear combination of original attributes, through which it can merge the tasks of understanding the distribution and semantics of features (e.g., clusters, outliers, general patterns) and sorting / filtering data into a unified exploration context. The visual analysis system complements this context by providing supporting components and rich interactions between them. The semantic axis is computationally efficient and can be used for large-scale data. However, the inherent linearity of our SemanticAxis may hinder its application in highly non-linear semantics analysis.

\acknowledgments{
The authors wish to thank A, B, and C. This work was supported in part by
a grant from XYZ (\# 12345-67890).}

\bibliographystyle{abbrv-doi}

\bibliography{template}
\end{document}